\documentclass[12pt]{article}

\usepackage{times}
\newenvironment{sciabstract}{%
\begin{quote} \bf}
{\end{quote}}

\usepackage{authblk}
\usepackage{graphicx}
\graphicspath{ {images} }
\usepackage[utf8]{inputenc}
\usepackage[T1]{fontenc}
\usepackage{amsmath}
\usepackage{amstext}
\usepackage{amsopn}
\usepackage{ amsbsy}
\usepackage{amsfonts}
\usepackage{amssymb}
\usepackage{graphicx}
\usepackage[nottoc]{tocbibind}
\usepackage{braket}
\usepackage{physics}
\usepackage{subcaption}
\usepackage{color}
\definecolor{prdblue}{rgb}{0.133,0.118,0.698}
\usepackage[colorlinks=true, pdfstartview=FitV, linkcolor=prdblue, citecolor= prdblue, urlcolor=prdblue]{hyperref}

\newcommand{\opp}{\textbf{p}}
\newcommand{\opE}{\textbf{H}}
\newcommand{\opx}{\textbf{x}}
\newcommand{\opA}{\textbf{A}}

\title{Theory of Neutrino Detection - Flavor Oscillations and Weak Values}

\author{{Yago} P. Porto-Silva$^{\ast}$, {Marcos} C. de Oliveira$^\dagger$
\\
\normalsize{ $^\ast$ $^\dagger$Instituto de F\'\i sica ``Gleb Wataghin'', Universidade Estadual de Campinas, Campinas, SP, Brazil}
\\
\normalsize{$^\ast$ Max-Planck-Institut  f\"ur  Kernphysik,  69117  Heidelberg,  Germany}
\\

\normalsize{E-mail: $^\ast$yporto@ifi.unicamp.br, $^\dagger$marcos@ifi.unicamp.br }
}

\date{}

\begin{document}
\maketitle

\begin{sciabstract}
  We show that, in the relativistic limit, the quantum theory of neutrino oscillations can be described through the theory of weak measurements with pre- and post-selection. The weak nature of neutrino detection allows simultaneous determination of flavor and energy without problems related to the collapse of the wavefunction. Together with post-selection, a non-trivial quantum interference emerges, allowing one to describe a flavor neutrino as one single particle, despite its superposition of masses. We write down the flavor equation of motion and calculate the flavor oscillation probability by showing precisely how a single neutrino interferes with itself.
\end{sciabstract}





\section*{Introduction}

Neutrino oscillations is an experimentally established phenomenon by which neutrinos undergo flavor transformations periodically as they propagate large enough distances \cite{Fukuda:1998mi,Ahmad:2002jz,Araki:2004mb}. It is described by a simple quantum mechanical model in which the flavor states are not eigenstates of the propagation Hamiltonian, but some linear combination of them, and, as a consequence, the flavor content changes with time and distance \cite{Pontecorvo:1957cp,Gribov:1968kq,Eliezer:1975ja,Fritzsch:1975rz,Bilenky:1975tb}.

Most of the literature treats the eigenstates of propagation as plane waves: states with definite mass, energy, and momentum. Besides, assumptions as same energies or same momenta for all the propagation eigenstates, also known as mass eigenstates, are usually made. All these assumptions are unphysical in the sense that they can violate coherence of mass eigenstates, energy-momentum conservation and are unable to describe space-time localized processes as neutrino production and detection \cite{Kayser:1981ye,Winter:1981kj,Giunti:2000kw}.
To overcome these difficulties many authors have proposed treatments based on neutrino quantum mechanical (QM) wave-packets \cite{Nussinov:1976uw,Giunti:1991ca,rich-PhysRevD.48.4318,Giunti:1997wq,Kiers:1995zj,Kiers:1997pe,Akhmedov:2009rb,Akhmedov:2017xxm,Akhmedov:2019iyt} and quantum field theory \cite{Kobzarev:1981ra,Giunti:1993se,Blasone:1995zc,Grimus:1996av,Beuthe:2001rc,Beuthe:2002ej,Giunti:2002xg,Akhmedov:2010ms,Kobach:2017osm}. Here, we explore neutrino oscillations at the level of QM wave-packets. 

In this paper, we shall work with Gaussian wave-packets\footnote{See \cite{Akhmedov:2010ms} for a discussion on the conditions under which a Gaussian envelope is a good approximation for an arbitrary wave-packet.}. The uncertainties in energy and momentum of a massive neutrino wave-packet come from the \textit{approximate} conservation of mean energies and momenta of all particles in the production and detection processes. When these uncertainties are large enough  so that one cannot, even in principle, resolve the masses, the produced and detected states can be written as a coherent superposition of mass eigenstates.
When neutrinos are detected, their flavor is revealed by the charged leptons produced in the interaction, and their energy and momentum can, in principle, be reconstructed by measuring energies and momenta of all other particles involved in the detection process. Even if all this information is inaccessible to the experimentalist, it is available to the particles in the detection, and this, by itself, configures a measurement. 
However, from QM, two \textit{incompatible} observables are being measured at the same time in this detection process: flavor and energy-momentum\footnote{Incompatible observables do not share the same set of eigenstates: the flavor eigenstates are certainly not the same as the energy-momentum eigenstates \cite{Sakurai:1167961}.}. 

The most critical consequence of measuring two incompatible observables at the same time is that they randomly mess up information about each other, being manifestly complementary. Therefore, in the same way that, in general, measuring momentum degrades the information about the position of quantum particles, the measurement of energy-momentum of the neutrino should disturb previous information about flavor, in particular, the flavor transitions during the propagation, and it would have been impossible to study neutrino oscillations.
The reason why the energy-momentum measurement does not prevent neutrino oscillations is, again, the large uncertainties in energy and momentum in the detection process. Indeed, this type of measurement, with large uncertainties, is an example of what is called in the literature \textit{weak measurements} \cite{non-commuting,GUDDER200518,Fuchs_1996} and their main feature is to disturb very little the quantum state of the system, not degrading the information about complementary observables\footnote{The concept of weak measurements has nothing to do with the concept of weak interactions in the Standard Model.}. Therefore, the same condition that allows neutrinos to be produced coherently also warrant flavor, energy, and momentum to be measured simultaneously by the detection particles.

Based on weak measurements plus post-selection \cite{PhysRevLett.60.1351,PhysRevA.41.11,Johansen_2004,Dressel_2014,shikano2011theory,Qin_2016}, \textit{i.e.}, the fact that usually only one flavor type is measured in the end, we develop the theory of neutrino detection and derive a relativistic quantum mechanics theory of particles described by a superposition of masses. We notice that flavor neutrinos obey Klein-Gordon continuity equation (ignoring spin) with momentum and energy described by \textit{weak values}\footnote{Weak values are the measured values when weak measurements with pre- and post-selection are performed.}. This is a {remarkable} connection between two completely independent developments in 20th-century quantum mechanics. From this formalism, the oscillation probability can be calculated without the conceptual issues usually encountered, such as the normalization problem \cite{Akhmedov:2019iyt,Akhmedov:2010ms}. 
{The weak features of neutrino oscillations were previously reported in the context of the spurious superluminal neutrino velocity \cite{Berry:2011ea,Tanimura:2011dq,Minakata:2012kg}. In this present work, however, we go beyond and promote weak values to the status of fundamental quantities in the description of neutrino oscillations. }

This paper develops as follows. Firstly we review the standard procedures for the description of neutrino wave-packets emission and detection and the resulting observed neutrino oscillations; Then we move on reviewing the von Neumann measurements and weak measurement regime. With all the fundamentals established, we develop the interpretation of neutrino oscillations under the  weak measurement regime, demonstrating how the neutrino probability current is described; and to finish we present our conclusions and perspectives.

\section*{Neutrino wave-packets}

In this section, we review the standard wave-packet formalism of neutrino oscillations in one dimension\footnote{This is a good approximation for cases in which the distance between neutrino source and detector is large compared to their size \cite{Akhmedov:2010ms}.}. We use natural units ($\hbar=c=1$) throughout the paper.

Consider a process at (average) coordinates $(t=0,x=0)$ that produces a neutrino of flavor $\alpha$ which propagates and is detected at $(T,L)$ with flavor $\beta$. Using a normalized Gaussian envelope, we can write the one-particle state of the neutrino produced at the origin:
\begin{equation} \label{initialstate}
    \ket{\nu_{\alpha}^P(0,0)}\equiv\ket{\nu_{\alpha}^P}= \sum_a U^{*}_{\alpha a} \ket{\nu_a^P}  =\sum_a U^{*}_{\alpha a}\int \frac{dp}{\sqrt{2 \pi}\sqrt{2 E_a(p)}} \hspace{0.05 cm}   \phi^P(p-p_a)  \ket{\nu_a(p)},
\end{equation}
with $E_a(p)=\sqrt{p^2+m_a^2}$ and
\begin{equation} \label{initial-gaussian}
   \int dp |\phi^P(p-p_a)|^2=1 \hspace{0.5cm} \rightarrow \hspace{0.5cm} \phi^P(p-p_a)=\frac{1}{(2 \pi \sigma_{pP}^2)^{\frac{1}{4}}} e^{-\frac{(p-p_a)^2}{4\sigma_{pP}^2}}.
\end{equation}
Here the flavor eigenstate $\ket{\nu_{\alpha}^P}$ is a superposition of mass eigenstate wave-packets, $\ket{\nu_a^P}$, with mass $m_a$, weightened by the complex-conjugated PMNS matrix elements, $U^{*}_{\alpha a}$. The mass eigenstates themselves are a superposition of energy and momentum eigenstates $\ket{\nu_a(p)}$. In the $x$-space, $\nu_a$ wave function, at time $t$, is given by \cite{Giunti:2007ry}:
\begin{equation} \label{space-time integral}
    \ket{\nu_a^P(x,t)}  =\int \frac{dp}{\sqrt{2 \pi}\sqrt{2 E_a(p)}} \hspace{0.05 cm}   \phi^P(p-p_a)  e^{-i E_a(p)t} e^{ipx}.
\end{equation}

The average momenta and momentum uncertainties of different mass eigenstates, $p_a$ and $\sigma_{pP}$, respectively, are determined by the kinematics and by the properties of the particles involved in the production ($P$) process. 
We assume all mass eigenstates are extremely relativistic, $p_a>>m_a$, so that we can approximate their average energies by \cite{Giunti:2007ry},
\begin{equation} \label{avE}
    \epsilon_a \approx E+\xi \frac{m_a^2}{2E},
\end{equation}
in which $E$ is the energy determined by the kinematics of the production process if neutrino masses are neglected and
\begin{equation}
    \frac{\xi}{2E}=\frac{\partial \epsilon_a}{\partial m_a^2} \bigg|_{m_a=0}
\end{equation}
is the coefficient of the first-order term if one expands $\epsilon_a=\sqrt{p_a^2+m_a^2}$ around $m_a = 0$. The corresponding momenta are
\begin{equation} \label{avP}
    p_a \approx E-(1-\xi)\frac{m_a^2}{2E}.
\end{equation}
For a given process, $\xi$ can be calculated from energy-momentum conservation up to order $\frac{m_a^2}{E^2}$.

The effective momentum-space uncertainty of the produced neutrino wave-packets $\sigma_{pP}$ is
\begin{equation} \label{sigmapP}
    \sigma_{pP} \sim \text{min}\{\delta_{pP},\delta_{eP}\}
\end{equation}
where $\delta_{pP}$ and $\delta_{eP}$ are, respectively, the momentum and energy uncertainties in the production process. In configuration-space, $\sigma_{xP}=\frac{1}{2 \sigma_{pP}}$.

The detection process, in the standard formalism, is considered by propagating the ket in (\ref{initialstate}) from the origin to $(T,L)$ and then projecting it on the state
\begin{equation} \label{finalstate}
    \ket{\nu_{\beta}^D}=\sum_a U^{*}_{\beta a}\int \frac{dp}{\sqrt{2 \pi}\sqrt{2 E_a(p)}} \hspace{0.05 cm}   \phi^D(p-p_a)  \ket{\nu_a(p)},
\end{equation}
with
\begin{equation} \label{final-gaussian}
    \int dp |\phi^D(p-p_a)|^2=1 \hspace{0.5cm} \rightarrow \hspace{0.5cm} \phi^D(p-p_a)=\frac{1}{(2 \pi \sigma_{pD}^2)^{\frac{1}{4}}} e^{-\frac{(p-p_a)^2}{4\sigma_{pD}^2}},
\end{equation}
which takes into account the effective momentum-space uncertainty $\sigma_{pD}$ of the detection ($D$) wave-packet, related to $\delta_{pD}$ and $\delta_{eD}$ in a similar way to (\ref{sigmapP}),
\begin{equation} \label{sigmapD}
    \sigma_{pD} \sim \text{min}\{\delta_{pD},\delta_{eD}\},
\end{equation}
and $\sigma_{xD}=\frac{1}{2 \sigma_{pD}}$. The average momentum $p_a$ seen in the detection process is determined by the kinematics of the production process\footnote{This constrain can be relaxed, see \cite{Akhmedov:2010ms}}. Notice that (\ref{initialstate}) and (\ref{finalstate}) are normalized independently.

Now, we compute
\begin{equation}
    A_{\alpha \beta}(L,T)=\bra{\nu_{\beta}^D}e^{-i \opE T+i \opp L}\ket{\nu_{\alpha}^P(0,0)}=\bra{\nu_{\beta}^D}\ket{\nu_{\alpha}^P(L,T)},
\end{equation}

\textit{i.e.}, the amplitude of probability of detection of neutrinos in state $\ket{\nu_{\beta}^D}$ when they were generated in state $\ket{\nu_{\alpha}^P(0,0)}$ after traveling the distance $L$ during the time interval $T$, being $\opE$ the Hamiltonian and $\opp$ the momentum operators. Using the condition $\bra{\nu_a(p)}\ket{\nu_a(p')}=(2 \pi) 2E_a(p) \delta(p-p')$:
\begin{equation} \label{pre-amp}
    A_{\alpha \beta}(L,T)=\frac{1}{(4 \pi^2 \sigma_{pP}^2\sigma_{pD}^2)^{\frac{1}{4}}} \sum_a U^{*}_{\alpha a}U_{\beta a} 
    \int dp \hspace{0.05 cm}     e^{-\frac{(p-p_a)^2}{4\sigma_{p}^2}} e^{-iE_a(p)T+ipL} .
\end{equation}

Let us consider sharply peaked Gaussian functions in momentum space\footnote{Dispersion due to different phase velocities is negligible \cite{Giunti:2002xg}.}, with average momentum and energy given by (\ref{avE}) and (\ref{avP}). In this context, the relativistic dispersion relation can be approximated by $E_a(p) \approx \epsilon_a + v_a(p-p_a)$, with
\begin{equation} \label{avV}
    v_a=\frac{\partial E_a(p)}{\partial p}\bigg|_{p=p_a}=\frac{p_a}{\epsilon_a}\approx 1-\frac{m_a^2}{2 E^2}.
\end{equation}
Thus,
\begin{equation} \label{pos-amp}
    A_{\alpha \beta}(L,T)= \sqrt{\frac{2 \sigma_{xP} \sigma_{xD}}{\sigma_x^2}} \sum_a U_{\alpha a}^{*} U_{\beta a}  \exp \bigg(-i \epsilon_{a}T+ip_{a}L-\frac{(L-v_{a}T)^2}{4\sigma_{x}^2} \bigg).
\end{equation}
In (\ref{pos-amp}), $\sigma_x$ is the effective size of the detection region\footnote{In the literature $\sigma_x$ is most commonly referred as the size of the wave-packet. Here we want to emphasize that it is related to the momentum resolution in the detection process.}, that takes into account space and time intervals in which the neutrino and all particles in the detection process are overlapped. Since the neutrino that reached the detection process carries information about the production process, $\sigma_x$ takes into account features of both production and detection, in a similar way to $\sigma_p$ - the effective resolution with which the detection process can measure momentum:
\begin{gather} \label{coh-uncertainties}
    \sigma_x^2=\sigma_{xP}^2+\sigma_{xD}^2 \hspace{0.5 cm} \text{and} \hspace{0.5 cm} \frac{1}{\sigma_p^2}=\frac{1}{\sigma_{pP}^2}+\frac{1}{\sigma_{pD}^2}.
\end{gather}
Both are related by $\sigma_{x}\sigma_{p}=\frac{1}{2}$. 

Squaring the amplitude in (\ref{pos-amp}) and integrating out the $T$ dependence, we obtain
\begin{multline} \label{prob1}
    P_{\alpha \beta}(L) = \frac{2 \sigma_{xP} \sigma_{xD}}{\sigma_x^2} \sum_{a,b} U_{\alpha a}^{*} U_{\beta a} U_{\alpha b} U_{\beta b}^{*}  e^{i(1-\xi)\frac{\Delta m^2_{ab}}{2E}L}  \\
    \times \int dT \exp \bigg[-\frac{(L-v_{a}T)^2+(L-v_{b}T)^2}{4\sigma_{x}^2} \bigg] e^{-i \xi \frac{\Delta m^2_{ab}}{2E}T}.
\end{multline} 
After integration, we substitute the expression in (\ref{avV}) for the velocity of relativistic mass eigenstates in the exponents, preserving terms of first order in $\frac{m_a^2}{E^2}$ (or one order higher if first order vanishes) and find
\begin{equation} \label{prob}
     P_{\alpha \beta}(L) = \frac{2\sqrt{2 \pi} \sigma_{xP} \sigma_{xD}}{\sigma_x} \sum_{a,b} \sqrt{\frac{2}{v_a^2+v_b^2}} U_{\alpha a}^{*} U_{\beta a} U_{\alpha b} U_{\beta b}^{*}  e^{-i\frac{\Delta m^2_{ab}L}{2E}} e^{-\big(\frac{L}{L_{coh}^{ab}}\big)^2} e^{-\frac{(\Delta \epsilon_{ab})^2}{8 \sigma_e^2}} ,
\end{equation}
with
\begin{equation} \label{decoh-terms}
    \Delta \epsilon_{ab}=\xi \frac{\Delta m_{ab}^2 L}{2E}, \hspace{0.5 cm} \text{and} \hspace{0.5 cm} L_{coh}^{ab}=\frac{4 \sqrt{2} E^2}{|\Delta m^2_{ab}|} \sigma_x,
\end{equation}
where $\sigma_e^2 \approx \frac{1}{2}(v_a^2+v_b^2)\sigma_p^2$. 

The probability in (\ref{prob}) is not normalized, and its magnitude is manifestly dependent on the sizes of produced and detected wave-packets and their overlap. Indeed, 
\begin{equation} \label{norm-problem}
    \sum_{\beta} P_{\alpha \beta}(L) = \frac{2\sqrt{2 \pi} \sigma_{xP} \sigma_{xD}}{\sigma_x} \sum_{a} \frac{|U_{\alpha a}|^2}{v_a} \approx \frac{2\sqrt{2 \pi} \sigma_{xP} \sigma_{xD}}{\sigma_x}.
\end{equation}
In addition, it is not dimensionless but has unit of length. Fixing it, therefore, is not a matter only of a constant factor, its calculation is conceptually incorrect. This is called the normalization problem and, to get rid of it, unitarity has to be imposed. This is rather unsatisfactory and a symptom that the formalism has consistency problems  \cite{Akhmedov:2019iyt,Akhmedov:2010ms}.

\begin{figure} 
\centering
\includegraphics[width=0.8\linewidth]{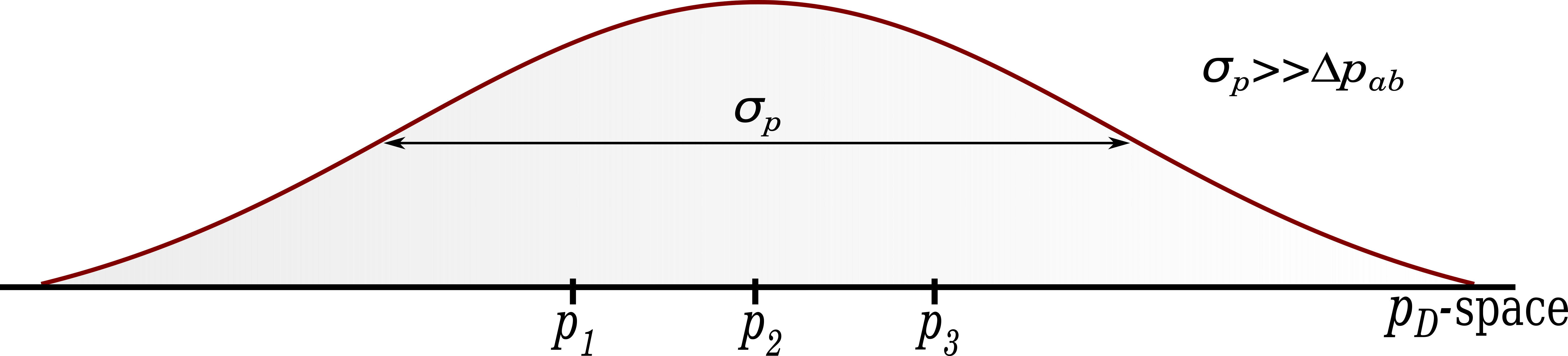} 
\caption{Illustration of the condition for coherent production and detection of neutrinos, $\Delta p_{ab}<<\sigma_p$, with $p_1$, $p_2$ and $p_3$ the mean momenta of the massive neutrino wave-packets given in (\ref{avP}).}
\label{coh-fig}
\end{figure}

The discussion about the physical meaning of the exponentials (\ref{prob}) can be found in many references \cite{Giunti:1991ca,Giunti:1997wq,Akhmedov:2019iyt,Giunti:2002xg,Giunti:2007ry}. Here we highlight:
\begin{itemize}
    \item The exponential $e^{-\big(\frac{L}{L_{coh}^{ab}}\big)^2}$ defines the coherence length, $L_{coh}^{ab}$, that is the effective distance after which mass eigenstates $\nu_a$ and $\nu_b$ lose coherence due to separation of their wave-packets. For $L<<L_{coh}^{ab}$ wave-packet separation is negligible.
    \item The term $e^{-\frac{(\Delta \epsilon_{ab})^2}{8 \sigma_e^2}}$ defines the conditions under which neutrinos are produced and detected coherently. In the limit, 
    \begin{equation} \label{coh-cond}
        \Delta \epsilon_{ab}<<\sigma_e, \hspace{1 cm}  \text{(Coherence Condition)}
    \end{equation}
    the conditions for coherent production and detection of the mass eigenstates $\nu_a$ and $\nu_b$ are set (see fig. \ref{coh-fig}). In the relativistic regime and in one dimension, (\ref{coh-cond}) is equivalent to $\Delta p_{ab}<<\sigma_p$ \cite{Akhmedov:2017xxm}.

\end{itemize}

Therefore, $L<<L_{coh}^{ab}$ and $\Delta E_{ab}<<\sigma_e$ are usually referred as the conditions for the observability of neutrino oscillations.

\section*{Weak measurements and weak values}

In this section, we formalize the concept of quantum measurement in the von Neumann regime and use it to distinguish between the strong (great disturbance, wavefunction collapse) and weak measurements (very little disturbance) \cite{non-commuting,GUDDER200518,Fuchs_1996,PhysRevLett.60.1351,PhysRevA.41.11,Johansen_2004}.

\subsection*{von Neumann measurements}

In the von Neumann measurement model \cite{VonNeumann:2302903, QuantumParadoxes}, the measuring device (or pointer) is a secondary quantum system with canonical variables $\opx_D$ and $\opp_D$ satisfying $[\opx_D,\opp_D]=i$, in natural units. The Hamiltonian that describes the interaction between the system and the device usually couples the observable of interest, that we call $\opA$, with some of the canonical variables, $\opx_D$, for example, so that the change in the conjugate variable, $\opp_D$, reveals information about $\opA$. The interaction Hamiltonian can be written as
\begin{equation} \label{H}
    \opE_{int}=-\delta(t-t_0) \opA \opx_D,
\end{equation}
where the delta function assures the interaction to happen for times only on the vicinity of $t_0$ while, at any other instant, the system evolves freely.
To illustrate how the pointer variable, $\opp_D$ in our example, acquires information about $\opA$, we compute its evolution at times close to $t_0$ in the Heisenberg picture:
\begin{equation} 
     \frac{d}{dt} \opp_D (t) = i[\opE_{int},\opp_D(t)]   = -i \delta(t-t_0) \opA(t)[\opx_D(t),\opp_D(t)] = \delta(t-t_0) \opA(t),
\end{equation}
then,
\begin{equation} \label{pointer}
    \opp_{D}(t>t_0)-\opp_{D}(t<t_0)=\int dt \delta(t-t_0) \opA(t) 
    =\opA(t_0).
\end{equation}
Therefore, the change in the pointer immediately after $t_0$ gives the information about the status of the observer of interest at $t_0$.

\subsection*{Statistics of the pointer variable}

Consider an ensemble defined by a system prepared in state $\ket{\psi_i}$ and measuring device in state $\ket{\phi}$. A system ensemble prepared in a specific initial state defines a preselected ensemble. We know that the effect of the measurement on the device is to change the status of its pointer variable $\opp_D$ proportionally to the system observable of interest $\opA$ according to (\ref{pointer}).
Starting from the initial state of the system plus measuring device, $\ket{\Omega_i}=\ket{\psi_i} \ket{\phi}$, we find that the impact of the measurement on this state is given by,
\begin{equation} \label{after}
    \ket{\Omega_f}=e^{-i \int \opE_{int} dt}\ket{\Omega_i}=e^{i \opA \opx_D} \ket{\psi_i} \ket{\phi}. 
\end{equation}
Projecting (\ref{after}) into the pointer variable space
\begin{equation} \label{translation}
    \bra{p_D}\ket{\Omega_f}
    =\sum_{a}\ket{a}\expval{a|\psi_i} \bra{p_D}e^{i a \opx_D}\ket{\phi}=\sum_{a}\ket{a}\expval{a|\psi_i} \phi(p_D-a),
\end{equation}
in which $\{\ket{a}\}$ are the eigenvectors of operator $\opA$, and $\phi(p_D-a)=\bra{p_D-a}\ket{\phi}$ is the shifted (by $a$) wavefunction, $\phi(p_D)$, due to the action of the translation operator $e^{i a \opx_D}$. 
The probability distribution of the pointer apparatus state after the measurement is given by the absolute square of (\ref{translation})
\begin{equation} \label{eigenvalues}
    P_f(p_D)=|\bra{p_D}\ket{\Omega_f}|^2=\sum_{a}|\expval{a|\psi_i}|^2 |\phi(p_D-a)|^2 .
\end{equation}
Remark that for the probability interpretation to hold we need a normalized pointer wavefunction,
\begin{equation} \label{norm-pointer}
    \expval{\phi | \phi}= \int dp_D |\phi(p_D)|^2 = 1.
\end{equation}

\begin{figure} 
\centering
\includegraphics[width=0.7\linewidth]{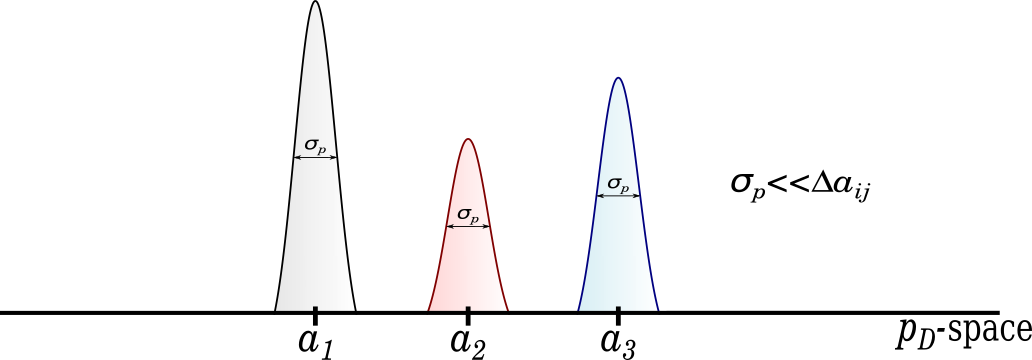} 
\caption{Illustration of a pointer wavefunction that \textit{can} resolve the spectrum of eigenvalues of the system observable $\opA$ in a strong measurement, $\sigma_{p} << \Delta a_{ij}$. The amplitude of the distributions are proportional to the probability amplitude of the system to be in the state $\ket{a_i}$. In the limiting case of $\sigma_p \rightarrow 0$, one recovers Born rule. Detector and system are fully entangled immediately after the measurement.}
\label{fig_strong}
\end{figure}

What we call strong or weak measurement depends very much on the spread of the apparatus wavefunction\footnote{It is common to model the pointer with a Gaussian wavefunction.} in the $p_D$-space, $\sigma_{p}$, relative to the separation, $\Delta a_{ij}$, of the eigenvalues, $\{a_i\}$, of the system. If the pointer can resolve the spectrum, in other words, if
\begin{equation}
    \sigma_{p} << \Delta a_{ij}, \hspace{1cm} \text{(Strong measurement)}
\end{equation}
it is called strong measurement, and is pictorially represented in fig. \ref{fig_strong}. 

\begin{figure} 
\includegraphics[width=1\linewidth]{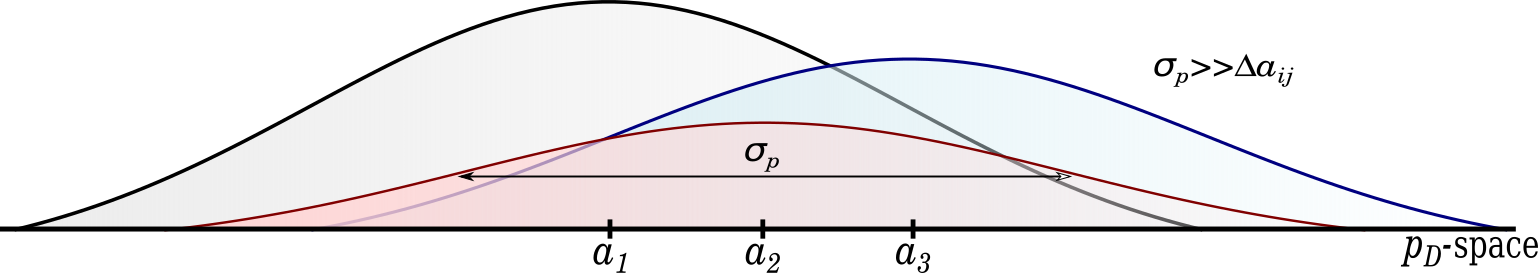} 
\caption{Illustration of a pointer wavefunction that \textit{cannot} resolve the spectrum of eigenvalues of the system observable $\opA$ in a weak measurement, $\sigma_{p} << \Delta a_{ij}$. The amplitude of the distributions are proportional to the probability amplitude of the system to be in the state $\ket{a_i}$. Because of the poor resolution, detector and system are not fully entangled and the system wavefunction is very little disturbed after the measurement.}
\label{fig_weak}
\end{figure}

In the opposite limit,
\begin{equation} \label{weakcond}
    \sigma_{p} >> \Delta a_{ij} \hspace{1cm} \text{(Weak measurement)}
\end{equation}
we say that the system is weakly measured by the apparatus, see fig. \ref{fig_weak}.

Weak measurements were first proposed as a way in which one can extract average state information without fully collapsing the system \cite{QuantumParadoxes}. In fact, due to the large uncertainty $\sigma_{p}$ in the apparatus wavefunction, after the measurement, the state of the apparatus is not strongly correlated or entangled with any of the states $\{\ket{a_i}\}$ of the system. This can be seen graphically in fig \ref{fig_weak}. For comparison, observe how the states of the system and apparatus are fully correlated after a strong measurement (resembling Born rule) in fig. \ref{fig_strong}.

A useful way of thinking about weak measurements is that the eigenvalues of $\opA$ are so close that the effect of the translation operator on the pointer is very small. More precisely, suppose our pointer just measured $a_1$ (went from zero to $a_1$ by means of $e^{i a_1 \opx_D}$) in fig. \ref{fig_weak}, to move its center to $a_2$ we operate with:
\begin{equation} \label{expansion}
    e^{i (a_2-a_1) \opx_D} =  1+ i \Delta a_{21} \opx_D - \frac{1}{2}(\Delta a_{21})^2  \opx_D^2 + ...
\end{equation}
with $\Delta a_{21}= a_2-a_1$. However, due to (\ref{weakcond}),
\begin{equation}
    \bra{\phi}(\Delta a_{21})^2  \opx_D^2 \ket{\phi}= (\Delta a_{21})^2 \sigma_{x}^2 \approx \frac{(\Delta a_{21})^2}{ \sigma_{p}^2}<<1.
\end{equation}
Hence, in case the measurement is weak, it is enough to use the expansion in (\ref{expansion}) up to first order.

Although strong and weak measurements are conceptually different, they are quantitatively equivalent with respect to expectation values \cite{Aharonov_2005}. In other words,
\begin{equation} \label{exp}
    \expval{\Omega_f|\opp_D|\Omega_f}=\expval{\opA}{\psi_i},
\end{equation}
independently of $\sigma_{p}$.

\subsection*{Pre- and post-selected ensembles}

If the expectation values of the observables of the system are the same independently of the measurement type (see (\ref{exp})), then one can judge unnecessary to talk explicitly about the weak nature of the neutrino energy-momentum measurement.
The problem is that, as we are going to see in next section, neutrino oscillation measurements are, in general, made in pre- and post-selected ensembles and the results of these measurements are not expectation values, but \textit{weak values}, a concept introduced in 1988 by Aharanov, Albert and Vaidman (AAV) \cite{PhysRevLett.60.1351,Dressel_2014,shikano2011theory,Qin_2016}.

\begin{figure} 
\includegraphics[width=1\linewidth]{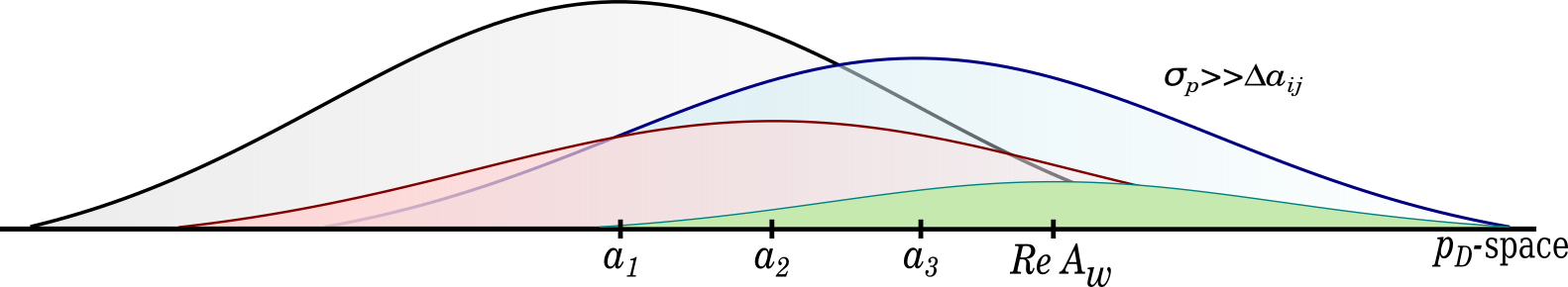} 
\caption{Illustration of the resulting pointer wavefunction (green), with mean given by the real part of the weak value, $\Re A_w$, when, together with weak measurements, there is post-selection of the state $\ket{\psi_f}$. Its amplitude is proportional to $\braket{\psi_f}{\psi_i}$. On purpose, $\Re A_w$ is shown outside the range of eigenvalues $a_i$, to highlight one of the most interesting properties of weak values.}
\label{fig_weak-value}
\end{figure}

To understand the concept of weak value, suppose that after the measurement described by the Hamiltonian in (\ref{H}) in the system ensemble $\Omega_{i}$, we focus only on the measurement outcomes from the system subensemble that ended up in some specific state $\ket{\psi_f}$. We name this subensemble $\Omega_{if}$.
We can find the statistics of the apparatus pointer variable by taking (\ref{after}) and applying to it the system conditional final state,
\begin{equation} \label{weak}
    \ket{\Omega_{if}}=\expval{\psi_f|\Omega_f}=\bra{\psi_f}e^{-i \int \opE_{int} dt}\ket{\Omega_i}=\bra{\psi_f}e^{i \opA \opx_D} \ket{\psi_i} \ket{\phi}.
\end{equation}
Now, we expand the exponential inside (\ref{weak}), 
\begin{equation} \label{second order}
     \ket{\Omega_{if}} \approx \expval{\psi_f|\psi_i}(1+i A_w \opx_D-\frac{1}{2}A^2_w \opx^2_D+...)\ket{\phi},
\end{equation}
where $A_w^n$ is called the n-th order weak value of $\opA$, $A_w^n\equiv\frac{\bra{\psi_f}\opA^n\ket{\psi_i}}{\expval{\psi_f|\psi_i}}$.
We use the hypothesis of weak measurements to argue that, for the apparatus, the action of the Hamiltonian in (\ref{H}) is just a small perturbation and truncate the expansion at first order in $\opx_D$, 
\begin{equation} \label{preamp}
     \bra{p_D}\ket{\Omega_{if}} \approx \expval{\psi_f|\psi_i} \bra{p_D}1+i A_w \opx_D\ket{\phi},
\end{equation}
using the approximation $1+i A_w \opx_D \approx e^{i A_w \opx_D}$, which acts as a translation operator in $p_D$-space
\begin{equation}
    \bra{p_D}1+i A_w \opx_D\ket{\phi} \approx \bra{p_D-A_w}\ket{\phi}.
\end{equation}
Assuming for the state of the pointer before the measurement,
\begin{equation} \label{gaussian_pointer}
    \expval{p_D|\phi} = \phi(p_D) \propto e^{-\frac{p_D^2}{4 \sigma_{p}^2}}, \hspace{1 cm} \text{with} \hspace{1 cm} \sigma_{x}\sigma_{p}=\frac{1}{2},
\end{equation}
and that $A_w$ is a complex number, $A_w=\Re A_w+i \Im A_w$, then, the probability distribution for the apparatus after the measurement is
\begin{equation} \label{weak-answer}
    |\expval{p_D-\Re A_w-i \Im A_w|\phi}|^2 \propto e^{\frac{(\Im A_w)^2}{2 \sigma_{p}^2}} e^{-\frac{(p_D-\Re A_w)^2}{2 \sigma_{p}^2}} \approx |\phi(p_D-\Re A_w)|^2,
\end{equation}
which is approximately the initial probablity distribution translated by $\Re A_w$ in $p_D$-space. Therefore, instead of moving the pointer to some eigenvalue, as in (\ref{eigenvalues}), a weak measurement with pre- and post-selection move the pointer to real part of the observable weak value, see fig. \ref{fig_weak-value}.

We usually think in eigenvalues as the only possible answers to single quantum measurements, but what this section teaches us is that with enough uncertainty in the detection and post-selection, an entirely new type of answer appears: the weak value, $A_w$. The weak value is considered a property of a single system under pre- and post-selection, revealed by a single measurement \cite{Vaidman_2017}. All the physical consequences of the interactions of the system under such circumstances depend on $A_w$.
Weak values can lie beyond the range of eigenvalues of $\opA$, so-called anomalous weak values, as illustrated in fig. \ref{fig_weak-value}, and, at its core, is a complicated quantum interference effect that is mathematically described by the concept of \textit{superoscillations} \cite{Aharonov_2011,Berry_2011}. 

Observe in fig. \ref{fig_weak-value} that, due to the uncertainty in the detection process, the eigenfunctions of the observable $\opA$ look like wave-packets with mean values $a_i$. Under general pre- and post-selection, \textit{i.e.} initial and final states are not restricted to be eigenvectors of observable $\opA$ - an effective wave-packet emerges with the mean value given by $\Re A_w$.  What comes in the following section can be already anticipated: if the total momentum uncertainty in the neutrino detection is large enough (see fig. \ref{coh-fig}) and we post-select a given flavor (see (\ref{finalstate})), then a kind of ``flavor wave-packet'' emerges with mean momentum given by a weak value\footnote{The same happens for energy.}.

\section*{Neutrino oscillations and the weak regime}

In this section we aim to construct a quantum theory of neutrino detection. Start by interpreting the detection region (see discussion before (\ref{coh-uncertainties})), as an apparatus (or pointer) that will measure neutrino energy and momentum with uncertainties $\sigma_e$ and $\sigma_p$, respectively. In the relativistic one-dimensional case, it is redundant to talk in terms of momentum and energy; then, in the following, we refer to momentum measurement. In the \textit{pointer interpretation}, (\ref{coh-cond}) must be understood in the same sense as (\ref{weakcond}):
\begin{equation} \label{weak-interp}
    \sigma_{p}>>\Delta p_{ab}. \hspace{1cm} \text{(Weak measurement)}
\end{equation}
As an apparatus, the detection region has conjugate variables $\opp_D$ and $\opx_D$ obeying $[\opx_D,\opp_D]=i$. In $p_D$-space, its wavefunction is given by the combination of the production and detection Gaussian envelopes in (\ref{initial-gaussian}) and (\ref{final-gaussian}), respectively\footnote{Actually, it is just after the measurement, $p_D \rightarrow p_D-p_a$, that $\phi_P$ and $\phi_D$ will be equal to the Gaussian evelopes in (\ref{initial-gaussian}) and (\ref{final-gaussian}).}:
\begin{equation} \label{nu-pointer}
    \braket{p_D}{\phi}=\phi(p_D)=\phi^P(p_D) \phi^D(p_D) \propto e^{-\frac{p_D^2}{4 \sigma_{pP}^2}} e^{-\frac{p_D^2}{4 \sigma_{pD}^2}} = e^{-\frac{p_D^2}{4 \sigma_{p}^2}},
\end{equation}
where we used (\ref{coh-uncertainties}).
Thus, we model the detection region as a Gaussian pointer with resolution $\sigma_p$, as in previous section. Here, $\sigma_p$ is the momentum resolution in neutrino detection, according to (\ref{coh-uncertainties}). 

We construct the Hamiltonian coupling the neutrino momentum, $\opp$, to the pointer conjugate variable, $\opx_D$ as\footnote{Notice that we are not imposing any kind of new interaction in the detection process, $\opE_{int}$, here, is just an artifact of calculation.}
\begin{equation} \label{approx_delta}
    \opE_{int}(t)=-\delta(t-T)  \opp \opx_D ,
\end{equation}
where $T$ is the average time of detection. According to this Hamiltonian, in analogy with (\ref{pointer}), after measurement (assuming initial value of $p_D$ is zero):
\begin{equation} 
    \opp_D (t>T)=\opp(t=T).
\end{equation}
In case the measurement is made for a neutrino mass eigenstate $\nu_a$, described by (\ref{initialstate}), we have, in Heisenberg picture (initial state $\ket{\nu_a^P}\ket{\phi}$),
\begin{equation} \label{pure}
    \bra{\nu_a^P}\bra{\phi}\opp_D(t>T)\ket{\phi}\ket{\nu_a^P}=\braket{\phi}{\phi}\bra{\nu_a^P}\opp(t=T)\ket{\nu_a^P}=p_a,
\end{equation}
with $\ket{\phi}$ the (normalized) state of the detection region. In other words, the detection region momentum distribution after the measurement ($t>T$) is clustered around $p_D=p_a$, given in (\ref{avP}), as expected. This is just telling us that the detection process behaves as if a wave-packet with mean momentum $p_a$ and uncertainty $\sigma_p$ just arrived. In our example, if $p_a$ is known, the corresponding energy, $\epsilon_a$, is also known.

Next subsection is devoted to the most general case of coherent detection of several mass eigenstates with post-selection (detection of a specific flavor). Weak values naturally appear.

\subsection*{Neutrino oscillations with pre- and post-selection}

\begin{figure} 
\includegraphics[width=1\linewidth]{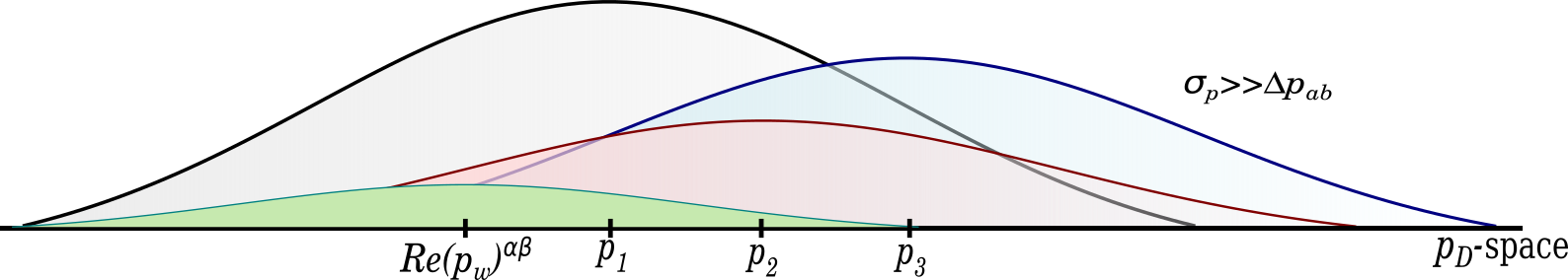} 
\caption{Illustration of the weak measurement features of neutrino oscillations. The detection of the flavor $\ket{\nu_\beta}$ constitutes post-selection. As usual, the resulting wave-packet (green) has mean given by $\Re p_w^{\alpha \beta}$ and we interpret it as the wave-packet of the detected $\nu_{\beta}$ or, generically, the ``flavor wave-packet''. Its amplitude is proportional to $\braket{\nu_{\beta}}{\nu_{\alpha}(L,T)}$.}
\label{nu-weak}
\end{figure}

In the case the neutrino is preselected in the state $\ket{\nu_{\alpha}^P}$, evolves freely to $\ket{\nu^P_{\alpha}(L,T)}$, until being detected and, consequently, post-selected in the state $\ket{\nu_{\beta}^D}$, we can write for the initial state
\begin{equation}
    \ket{\Omega_{\alpha}}=\ket{\nu_{\alpha}^P(L,T)} \ket{\phi}.
\end{equation}
 In analogy with (\ref{weak}),
\begin{eqnarray} 
    \ket{\Omega_{\alpha \beta}(L,T)}&=&\expval{\nu_{\beta}^D|\Omega_{\alpha}}=\bra{\nu_{\beta}^D}e^{-i \int \opE_{int} dt} \ket{\nu_{\alpha}^P(L,T)} \ket{\phi}  \nonumber\\&=& \bra{\nu_{\beta}^D}e^{i \opp \opx_D} \ket{\nu_{\alpha}^P(L,T)} \ket{\phi}.
\end{eqnarray}
Using the weak measurement hypothesis (\ref{weak-interp}),
\begin{eqnarray} \label{nu-second-order}
         \ket{\Omega_{\alpha \beta}(L,T)} &\approx& \expval{\nu^D_{\beta}|\nu^P_{\alpha}(L,T)} \bigg(1+i p^{\alpha \beta}_w \opx_D \bigg)\ket{\phi} \nonumber\\ &\approx & \expval{\nu^D_{\beta}|\nu^P_{\alpha}(L,T)} \ket{\phi(p_D-\Re \{p^{\alpha \beta}_w\})},
\end{eqnarray}
where $p^{\alpha \beta}_w$ is also a function of $L$ and $T$, given by
\begin{equation} \label{weak-mom}
    p^{\alpha \beta}_w(L,T)=\frac{\bra{\nu^D_{\beta}} \opp \ket{\nu^P_{\alpha}(L,T)}}{\expval{\nu^D_{\beta}|\nu^P_{\alpha}(L,T)}}.
\end{equation}
Hence, the neutrino momentum measured by the particles in the detection region at average coordinates $(T,L)$ is given by $p_D=\Re \{p^{\alpha \beta}_w(L,T)\}$. Analogously, the energy is the real part of
\begin{equation}
    \epsilon^{\alpha \beta}_w(L,T)=\frac{\bra{\nu^D_{\beta}} \opE \ket{\nu^P_{\alpha}(L,T)}}{\expval{\nu^D_{\beta}|\nu^P_{\alpha}(L,T)}}.
\end{equation}
Notice that the flavor neutrino behaves as a single particle wave-packet with average energies and momenta $ \Re \epsilon^{\alpha \beta}_w$ and $\Re p^{\alpha \beta}_w$ at $(T,L)$, see fig. \ref{nu-weak}, in the same sense the massive wave-packets have averages $\epsilon_a$ and $p_a$. As explained after (\ref{weak-answer}), the detection process effectively sees a ``flavor wave-packet'' as a consequence of large uncertainties and post-selection. In this \textit{specific} context, flavor neutrinos can be considered particles with their own wave-packets.

\subsection*{Normalization and probability current}

In this subsection, we work with one massive neutrino $\nu_a$, mass $m_a$, and explain how to write its wavefunction, probability density and current satisfying the pointer interpretation. Any other massive particles, such as electrons or muons, would have the same treatment. In the next subsection we mix the massive neutrinos and find an analogous treatment for flavor neutrinos.

Eqs. (\ref{norm-pointer}) and (\ref{nu-pointer}) imply that for our interpretation of detection region as a pointer, production and detection Gaussian envelopes should not be normalized separately but in a correlated manner,
\begin{equation} \label{corr-norm}
        \int dp \hspace{0.1cm} |\phi(p-p_a)|^2 = \int dp \hspace{0.1cm} |\phi^P(p-p_a)|^2|\phi^D(p-p_a)|^2=1,
\end{equation}
and therefore
\begin{equation} \label{corr-gaussian}
    \phi(p-p_a)=\frac{1}{(2 \pi \sigma_p^2)^{\frac{1}{4}}} e^{\frac{(p-p_a)^2}{4 \sigma_p^2}}.
\end{equation}
This has a simple interpretation -- the final momentum distribution, $\phi$, represents the detection of \textit{one} particle independently of how exactly $\phi^P$ and $\phi^D$ overlap\footnote{\label{double-slit} This is a feature that every measurement formalism in QM must take care of. Think of the double-slit experiment, there is a particle propagating as a wave, this wave partially reflects in the wall, partially goes through one slit or the other, but when we get to measure the particle on the other side of the wall, it is the \textit{entire} particle on the spot. It is a sharp "click" of the detector, and the intensity of the click does not depend on factors such as how the wave overlaps with the slit. These factors become important in the many-particle cases, for predicting the rate of clicks at a specific location.}. The coordinate space wavefunction at time $T$ and position $L$ for this particle is, in analogy with (\ref{space-time integral}), the space-time integral of $\phi(p-p_a)$:
\begin{equation}\label{good-vibes}
          A_{a}(L,T)=\bra{\nu_a^D}\ket{\nu_a^P(L,T)}=\int \frac{dp}{\sqrt{2 \pi} \sqrt{E_a(p)}} \hspace{0.05 cm}     \phi(p-p_a) e^{-iE_a(p)T+ipL} .
\end{equation} 
Hence, instead of propagating the produced state to the detected state, we bring produced and detected states together and propagate them as one particle\footnote{We write $A_a=\bra{\nu_a^D}\ket{\nu_a^P(L,T)}$ to keep track that $\phi(p-p_a)=\phi^P(p-p_a)\phi^D(p-p_a)$, therefore the integral is a ``correlated'' inner product. In a sense, this is a time symmetric formulation \cite{PhysRev.134.B1410,Reznik_1995,article}.}. For sharply peaked wave-packets,
\begin{eqnarray} \label{sharp-good-vibes}
    A_{a}(L,T) &\approx& \int \frac{dp}{\sqrt{2 \pi}\sqrt{2 \epsilon_a}} \phi(p-p_a) e^{-i \epsilon_a T} e^{-i v_a(p-p_a) T} e^{i p L} \nonumber\\ &=& \frac{1}{\sqrt{2 \pi}\sqrt{2 \epsilon_a}}\bigg(\frac{2 \pi}{\sigma_x^2}\bigg)^{\frac{1}{4}} \exp \bigg(-i \epsilon_{a}T+ip_{a}L-\frac{(L-v_{a}T)^2}{4\sigma_{x}^2} \bigg).
\end{eqnarray}

Relativistic particles with defined masses, such as $\nu_a$, obey Klein-Gordon equation\footnote{As matter of fact, $\nu_a$ obeys Dirac equation. Klein-Gordon density and currents are approximations of their respective Dirac counterparts when spinor degrees of freedom are ignored. The calculations in the following can be reproduced without ignoring the spinors by using Gordon decomposition \cite{sakurai1987advanced}.}. From (\ref{sharp-good-vibes}), the Klein-Gordon current for an arbitrary particle produced as $\ket{\nu_a^P}$ and detected as $\ket{\nu_a^D}$ after propagating a distance $L$ during some time $T$ is given by
\begin{equation} \label{mass-current}
    J_a(L,T)=2 p_a |A_{a}(L,T)|^2.
\end{equation}
Together with the probability density, $\rho_a(L,T)=2 \epsilon_a |A_{a}(L,T)|^2$, $J_a(L,T)$ satisfies the Klein-Gordon continuity equation,
\begin{equation} \label{mass-continuity}
    \frac{\partial}{\partial T}\rho_a(L,T) + \frac{\partial}{\partial L} J_a(L,T)=0.
\end{equation}
What we call probability density is actually a number (of particles) density. The way $A_a(L,T)$ is normalized in (\ref{good-vibes}), however, is so that there is only one massive neutrino $\nu_a$ in the whole space at any given time:
\begin{equation} \label{impose}
    \int dL \hspace{0.1cm} \rho_a(L,T)= 1.
\end{equation}
Therefore, $A_a(L,T)$ satisfies two constraints, that we summarize:
\begin{enumerate}
    \item For the pointer interpretation of the detection process, we need a correlated normalization of produced and detected envelopes in (\ref{corr-norm}).
    \item For the probability interpretation of the neutrino wavefunction in a relativistic theory, we impose that the number of neutrinos in all space at any given time is one, (\ref{impose}).
\end{enumerate}
Note that, with such a convention, $\rho_a$ has dimension of  1/\textit{length} as it should be in a one-particle theory and $J_a \propto$ 1/\textit{time}. Therefore, integrating $J_a$ for the whole time of the experiment should give us the probability that, after the detection or "click", the detector will register a particle of index $a$:
\begin{equation} \label{impose2}
    P_a(L)=\int dT \hspace{0.1cm} J_a(L,T)= 1.
\end{equation}

For the case of particles with superposition of masses, as flavor neutrinos, it is not straightforward to find the corresponding (\ref{mass-current}) and (\ref{mass-continuity}) \cite{Zralek:1998rp,Ancochea:1996wu,Blasone:2001qa,Blasone:2002wp}. But, in the next section, we use the interpretation of ``flavor wave-packets'', that comes from weak measurement and post-selection, to guess the form of their probability current. Presumably, after time integration, it might give us the expression for the flavor oscillation probability.

\subsection*{Neutrino oscillation Probability}

Taking seriously the idea that the flavor neutrinos can be considered particles described by the ``flavor wave-packets'', we guess the form of their probability current by analogy with eq. \ref{mass-current}:  
\begin{equation} \label{flavor-current}
    J_{\alpha \beta}(L,T)=2 \Re\{p_w^{\alpha \beta}\} |A_{\alpha \beta}(L,T)|^2,
\end{equation}
with $A_{\alpha \beta}(L,T)=\sum_a U_{\alpha a}^* U_{\beta a} A_a(L,T)$ and $\Re\{p_w^{\alpha \beta}\}$ in place of the average momentum of the wave-packet. It can be shown that $J_{\alpha \beta}$ obeys a continuity equation of the form:
\begin{equation} \label{flavor-cont}
    \sum_{\beta} \bigg( \frac{\partial}{\partial T}\rho_{\alpha \beta}(L,T) + \frac{\partial}{\partial L} J_{\alpha \beta}(L,T) \bigg)=0, 
\end{equation}
with $\rho_{\alpha \beta}(L,T)=2 \Re\{\epsilon_w^{\alpha \beta}\} |A_{\alpha \beta}(L,T)|^2$. Indeed, (\ref{flavor-current}) and (\ref{flavor-cont}) can be derived from manipulating Klein-Gordon equation without ever referring to weak measurements. Thus, weak values spontaneously appear and highlight the underlying weak regime in the physics of mixed particles\footnote{This is the content of a future paper.}. Different from (\ref{mass-current}), (\ref{flavor-current}) describes a probability that is not conserved, in general, due to flavor transformations.

The time-independent flavor oscillation probability is defined as probability current integrated over the area of the detector for the entire time duration of the experiment,
\begin{equation}
    P_{\alpha \beta}(\vec{L})=\int dT \int_S d A \hspace{0.1 cm} J_{\alpha \beta}(\vec{L},T).
\end{equation}
Since we are working in just one dimension, this integral simplifies to
\begin{equation} \label{int-current}
    P_{\alpha \beta}(L)=\int dT \hspace{0.1 cm} J_{\alpha \beta}(L,T).
\end{equation}
This is equivalent to (\ref{impose2}) in the context of mixed particles, it gives the probability that the detector will register index $\beta$ after the "click". Substituting (\ref{flavor-current}) into (\ref{int-current}), we have
\begin{eqnarray} \label{prob2}
           P_{\alpha \beta}(L)&=&\int dT \hspace{0.1 cm} 2 \Re\{p_w^{\alpha \beta}\} |A_{\alpha \beta}(L,T)|^2 \nonumber \\  
           &=&2 \Re   \sum_{a,b} U^*_{\alpha a} U_{\beta a} U_{\alpha b} U^*_{\beta b} \int dT \expval{\nu_{b}^P(L,T)|\nu_{b}^D} \bra{\nu^D_{a}} \opp \ket{\nu^P_{a}(L,T)}. \label{T-int}
\end{eqnarray}
Now, $\bra{\nu^D_{a}} \ket{\nu^P_{a}(L,T)} \approx A_a(L,T)$ as given by (\ref{sharp-good-vibes}). For the second term in (\ref{prob2}),
\begin{eqnarray}
    \bra{\nu^D_{a}} \opp \ket{\nu^P_{a}(L,T)} &\approx& \int \frac{dp}{\sqrt{2 \pi}\sqrt{2 \epsilon_a}} \hspace{0.1 cm} p \hspace{0.1 cm} \phi(p-p_a) e^{-i \epsilon_a T} e^{-i v_a(p-p_a) T} e^{i p L} \nonumber\\  &=& \bigg( p_a+2i\frac{L-v_aT}{4 \sigma_x^2} \bigg)  \frac{1}{\sqrt{2 \pi}\sqrt{2 \epsilon_a}}\bigg(\frac{2 \pi}{\sigma_x^2}\bigg)^{\frac{1}{4}} \nonumber\\&&\times\exp \bigg(-i \epsilon_{a}T+ip_{a}L-\frac{(L-v_{a}T)^2}{4\sigma_{x}^2} \bigg) ,
\end{eqnarray}
or
\begin{equation}
    \bra{\nu^D_{a}} \opp \ket{\nu^P_{a}(L,T)} \approx \bigg( p_a+2i\frac{L-v_aT}{4 \sigma_x^2} \bigg) A_a(L,T).
\end{equation}
Back to (\ref{T-int}), we have 
\begin{eqnarray} \label{right-int}
         P_{\alpha \beta}(L)&=&\frac{1}{ \pi} \bigg(\frac{2 \pi}{\sigma_x^2}\bigg)^{\frac{1}{2}}  \Re \bigg\{  \sum_{a,b} U^*_{\alpha a} U_{\beta a} U_{\alpha b} U^*_{\beta b} p_a  \frac{1}{\sqrt{2\epsilon_a}}\frac{1}{\sqrt{2\epsilon_b}}e^{i(1-\xi)\frac{\Delta m^2_{ab}}{2E}L}  \nonumber \\
         &&\times\int dT \exp \bigg[-\frac{(L-v_{a}T)^2+(L-v_{b}T)^2}{4\sigma_{x}^2} \bigg] e^{-i \xi \frac{\Delta m^2_{ab}}{2E}T} \bigg\},
\end{eqnarray}
in which relations from (\ref{avE}) and (\ref{avP}) have been used\footnote{We neglect the integral involving the term $2i\frac{L-v_aT}{4 \sigma_x^2}$.}. After integration,
\begin{equation} \label{quasi-prob}
     P_{\alpha \beta}(L) \approx   \Re  \bigg\{ \sum_{a,b} U^*_{\alpha a} U_{\beta a} U_{\alpha b} U^*_{\beta b} \sqrt{\frac{2}{v_a^2+v_b^2}}p_a  \frac{1}{\sqrt{\epsilon_a}}\frac{1}{\sqrt{\epsilon_b}}e^{-i\frac{\Delta m^2_{ab}L}{2E}} e^{-\big(\frac{L}{L_{coh}^{ab}}\big)^2} e^{-\frac{(\Delta \epsilon_{ab})^2}{8 \sigma_e^2}} \bigg\}.
\end{equation}
According to (\ref{quasi-prob}), $P_{\alpha \beta}$ is dimensionless, $\sum_{\beta} P_{\alpha \beta}(L) = 1$ and there is no normalization problem (see (\ref{norm-problem})). When the dependence of $p_a$, $\epsilon_a$ and $v_a$ on the index $a$ is negligible,
\begin{equation}
    \sqrt{\frac{2}{v_a^2+v_b^2}}p_a  \frac{1}{\sqrt{\epsilon_a}}\frac{1}{\sqrt{\epsilon_b}} \approx 1.
\end{equation}
Moreover, the result of the summation in (\ref{quasi-prob}) is real, and therefore
\begin{equation} \label{final-prob}
      P_{\alpha \beta}(L) \approx  \sum_{a,b} U^*_{\alpha a} U_{\beta a} U_{\alpha b} U^*_{\beta b} e^{-i\frac{\Delta m^2_{ab}L}{2E}} e^{-\big(\frac{L}{L_{coh}^{ab}}\big)^2} e^{-\frac{(\Delta \epsilon_{ab})^2}{8 \sigma_e^2}}.
\end{equation}
Hence, the probability that the detector will register index $\beta$ depends on the distance $L$ between the source and the detector as well as the energy $E$, computed for the case of massless neutrinos (see (\ref{avE})). This formula, which is the basis of all neutrino oscillations phenomenology, has an oscillatory behavior, and its frequency is proportional to the squared mass differences, $\Delta m^2_{ab}$. Here we derived it in a different route, appealing to the features of the detection process and the most basic principles of QM: the  complementarity, through the uncertainty principle and the measurement postulate. It is not exaggerated to say that the present theory addresses each of the issues that a complete quantum theory should present. 

Notice that, in case $L>>L_{coh}$ or $\Delta \epsilon_{ab} >> \sigma_e$, the mass eigenstates can be resolved and the measurement is strong by definition, implying the detection would collapse the wavefunction of the single flavor neutrino to one of its mass eigenstates. In such a situation, the interference of the neutrino with itself is inaccessible, and the detection probability should be described by a statistical mixture of the mass eigenstates destroying the oscillatory pattern of (\ref{final-prob}).

It is important to say that, in particle physics, the quantum \textit{one} particle treatment is not in general applicable. Elementary interactions involve the creation and annihilation of particles, and quantum field theory must be applied. In special cases of ultra-relativistic neutrinos in the laboratory frame, with sharply peaked momentum distributions and scattering amplitudes that are insensitive to the absolute masses, $m_a$, the many contributions to the detection rate can be factored, and a definition of oscillation probability makes sense \cite{Giunti:2002xg,Akhmedov:2010ms,Giunti:2007ry}. In such situations, that happens to be the one of most practical interest, neutrino flavor states are useful approximations, and the theory of neutrino oscillations presents us with all the richness of QM.

\section*{Conclusion}

This work revolves around the idea that, in the relativistic limit, the features of neutrino detection in
oscillation experiments are well described by the theory of weak measurements with pre- and post-selection. From this simple observation, everything else follows.

On the one hand, the weak nature of the phenomenon enables us to reconcile the fact that energy-momentum and flavor are measured simultaneously during the neutrino detection even though, in the quantum mechanical model of neutrino oscillations, they are incompatible observables. On the other hand, in analogy with the concept of weak values, we can describe flavor neutrinos as single particles with their own wave-packets. The ``flavor wave-packet'' is, then, the consequence of a highly non-trivial quantum interference effect that happens due to quantum uncertainties and post-selection. In mathematical physics, this interference effect is called \textit{superoscillations}.

With the wave-packet description of flavor neutrinos, we can treat them as single particles in despite their superposition of masses. As relativistic particles, they obey a specific type of Klein-Gordon continuity equation (by ignoring spin) and, therefore, they have an associated probability current. From the probability current, it is straightforward to calculate the time-independent flavor oscillation probability. The connection between the Klein-Gordon equation of motion and weak-values -- two completely independent developments in quantum mechanics -- is one of the main results of this paper.

The previous quantum mechanical treatments of neutrinos oscillations are unsatisfactory due to problems that come either from the use of plane waves or wave-packet treatments that do not address some of basic principles deep enough. The treatment in this present paper tells the narrative, step by step, of how a single neutrino interferes with itself. 

\section*{Acknowledgements}
This study was financed in part by the Coordenação de Aperfeiçoamento de Pessoal de Nível Superior - Brasil (CAPES) - Finance Code 001. YPPS acknowledges support from FAPESP funding Grants No. 2014/19164-
453 6, No. 2017/05515-0 and No. 2019/22961-9. MCO also acknowledges support from CNPq. YPPS is thankful to O. L. G. Peres, M. E. Chaves, A. Y. Smirnov and E. Akhmedov for enlightening discussions on wave-packets and neutrino oscillations.

\bibliographystyle{unsrt}
\bibliography{refs.bib}
\end{document}